\let\TeXyear\year
\documentclass{jaa}
\let\setyear\year
\let\year\TeXyear
\usepackage{natbib}
\usepackage{orcidlink}

\usepackage{graphicx}

\setyear{2024}

\begin{document}\sloppy

%%paper title
%%For line breaks \\ can be used within title
\title{Exoplanet Detection : A Detailed Analysis} 

%%author names are separated by comma (,)
%%use \and before the last author name
%%use a * along with the number separated by comma
%% for the  author for correspondence
%%\textsuperscript{number} is used for affiliation
%%\affilOne, \affilTwo etc., upto \affilTwentyfive is possible
%%Please note the first letter after \affil is capitalised in the command
%%

\author{Mahima Kaushik\textsuperscript{1,*} \orcidlink{0000-0003-2650-6758}, Aditee Mattoo\textsuperscript{1} and Dr. Ritesh Rastogi\textsuperscript{2}}
\affilOne{\textsuperscript{1}Department of M.Tech Integrated (CSE), Noida Institute of Engineering and Technology, Greater Noida, 201310, U.P, India\\}
% \affilTwo{\textsuperscript{2}Department of M.Tech Integrated (CSE), Noida Institute of Engineering and Technology, Greater Noida, 201310, U.P, India\\}
\affilThree{\textsuperscript{2}Department of CS, Noida Institute of Engineering and Technology, Greater Noida, 201310, U.P, India.}

%%escape two column mode for title, affiliation and abstract
%%by giving \twocolumn command as shown

\twocolumn[{

\maketitle

%%include \corres to print the corresponding author Email id
\corres{mahimakaushik333@gmail.com}

%%include \msinfo for
%%manuscript information such as
%%received, revised and accepted dates
%%
% \msinfo{}{}

%%abstract
\begin{abstract}
    The exoplanet detection is the most exciting and challenging field of astronomy. 
    The discovery of many exoplanets has revolutionized our understanding of the formation and 
    evolution of planetary systems and has showed new ways to search for extra terrestrial life. 
    In recent years, some primary methods of exoplanet detection like transit, radial velocity, gravitational 
    microlensing, direct imaging and astrometry have played a important role for the discovery of exoplanets. 
    In this paper we explored detection methodologies with all the implications and analytics of comparison between 
    them. 
    Here we also discussed on different machine learning algorithms for exoplanet detection and visualization. 
    Finally, concluded with the significant discoveries made by some missions and their implications on our 
    understanding for the properties, environmental conditions and importance of exoplanets in the universe.
  
\end{abstract}

%%insert keywords separated by 3 hyphens using \keywords{words}

}]
%%close the twocolumn escape here

%%include \doinum{number}for the DOI number in the header
%%include \volnum{number} for the volume number in the header
%%include \year{yyyy} for  year of publication in the header
%%include \pgrange{num--num} page range of article in the header
%%include \artcitid{num} for the article citation id
%%include \lp to print last page of the article
%%include \setcounter{page}{pagenum} for the exact starting page of the article

\doinum{10.48550/arXiv.2404.09143}
\artcitid{\#\#\#\#}
\volnum{000}
\setyear{2024}
\pgrange{1-10}
\setcounter{page}{1}
\lp{10}

\section{Introduction}
Since the discovery of exoplanets began, they have continually surprised us with the hidden 
secrets of the cosmos. Here on earth, wealth is often associated with the possession of precious 
rocks and metals. But what about worlds made of these rocks and metals? For instance, 55 Cancri e 
is an exoplanet made of diamonds and orbits a sun-like star. It was discovered by McArthur \cite{mcarthur2004detection} using 
the radial velocity method by observing the spectrum of one of the parent stars in a binary system 
located 40 light-years from the planet Earth. Exoplanets are poised to become a significant part of 
human civilization in various aspects, such as energy sources, colonization of other planets, transportation, 
space exploration, employment, and many others. Since the first exoplanet was discovered, astronomers have 
been working tirelessly to discover other extrasolar planets. Numerous missions have been conducted by 
several space agencies to study exoplanets, and the results have been truly fulfilling. Till date, 
approximately 5500+ exoplanets have been discovered \cite{akeson2013nasa}, with many of them undergoing detailed studies.
\begin{figure*}
  \centering
  \includegraphics[width=1\textwidth]{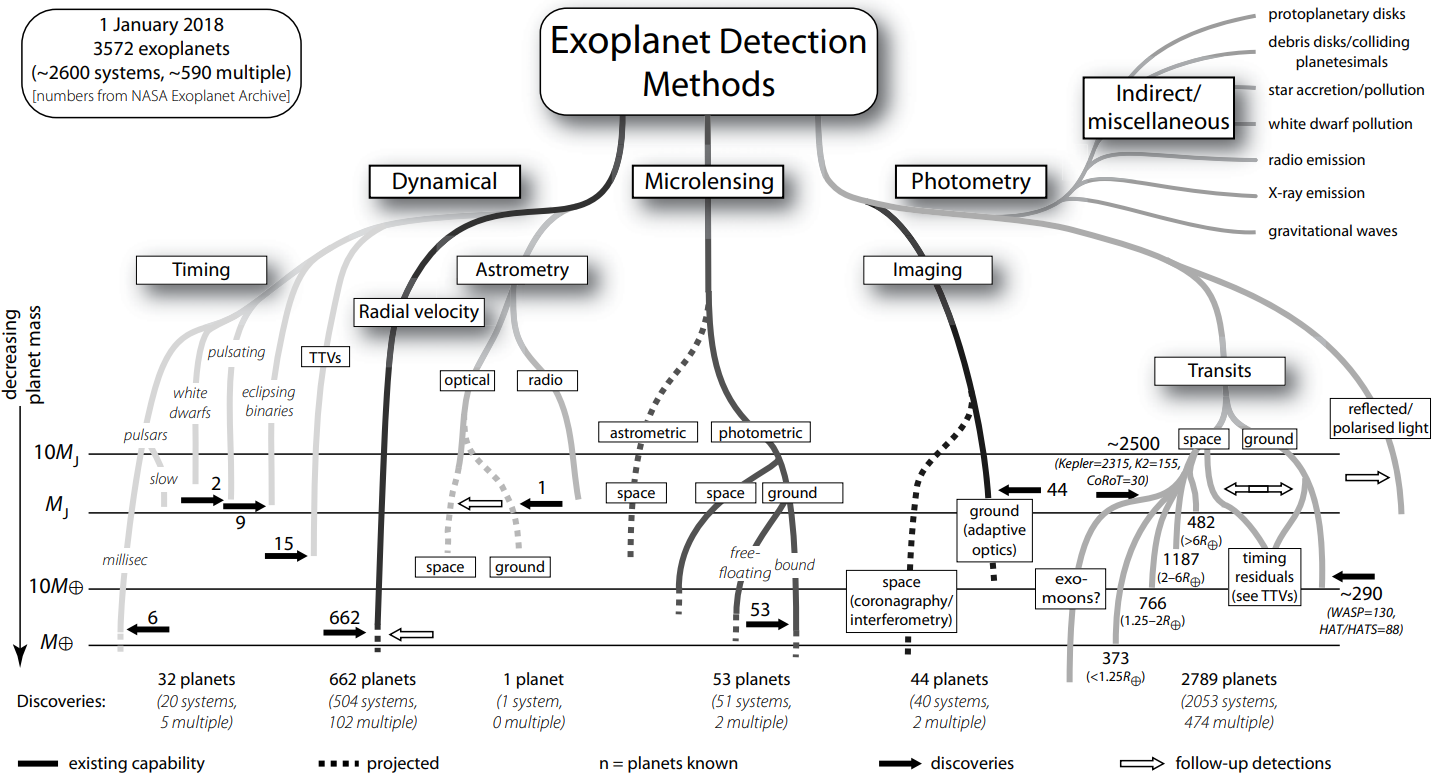}
  \caption{\scriptsize{Exoplanet detection methods, shown on a logarithmic mass scale, 
  show current and upcoming measuring possibilities. Open arrows show previous evaluations, 
  whereas solid arrows represent discoveries. Data collected from the NASA Exoplanet Archive (2018) \cite{perryman2018exoplanet}.}}
  \label{fig:exoplanets1}
\end{figure*}
 \vspace{4px}
Exoplanets come in various types such as earth-size, neptune-size, super-earth,  
jupiter-size, sub-earth, super-jupiters, brown dwarfs, and stellar companions. There are also 
various techniques for detecting exoplanets \cite{wei2018survey}, including radial velocity, transit method, astrometry, 
direct imaging, and gravitational microlensing. Each method offers distinct perspectives, strengths, 
limitations, and other details, often represented in tabular and graphical forms. By analyzing the 
light that passes through the atmosphere of a planet, scientists can understand its composition. Some 
exoplanets have been found to have atmospheres containing elements such as carbon dioxide, methane, and 
even water vapor \cite{perryman2018exoplanet}. If we compare a piece of wood and a piece of diamond to see which one is more 
expensive, the correct answer is wood. This is because, so far, earth is the only place in the universe 
where wood is produced, while diamonds are present on different exoplanets like 55 Cancri e. Gliese 667Cc 
is a super-earth exoplanet that lies 22 light-years away from earth. It's a potentially habitable super-earth 
with a mass of 3.8 times that of earth. Humans are still in the race of hunting for those habitable exoplanets, 
on which we can grow more plants and expand human colonization beyond Earth.

\vspace{4px}
51 Pegasi b, the first exoplanet, orbiting around a star similar to sun, was found using the radial 
velocity approach by Mayor \& Queloz. Several space agencies are running missions specifically for 
exoplanet detection using different detection methods \cite{mediumshafidetection1}. These missions include hubble, spitzer, corot, kepler/K2, 
gaia, TESS, CHEOPs, webb, and many more, which continuously monitor the universe, and astronomers keep detecting 
exoplanets with their help. In the future, there will be missions for specific exoplanets to understand 
their properties in detail and their environmental conditions. However, as of now, it's hard to travel to 
even nearby exoplanets within a lifetime. Further more, introduced space missions briefly, including details 
about data formats and other details. Astronomers use various machine learning \cite{kumari2023identification} and data cleaning algorithms 
to work on the data that comes from these missions. Such algorithms include the BLS algorithm, light curve 
analyzer, CNN, and other algorithms. all exoplanets are formed by normal matter but some scientists assumes 
that some exoplanets could me formed through dark matter \cite{bai2023dark}. Introduction about exoplanets with 
there basic details are going to be dicsussed in the next section.

\section{The exoplanets}
Extrasolar planets, also referred to as exoplanets, are planets that circle stars outside of our solar 
system. From the first exoplanet to be discovered in the early 1990s to the current booming field with 
thousands of confirmed planets, the growth of exoplanet investigation has been incredible \cite{wiki:exosearchprojects}. Section 2 discusses 
the crucial roles that various detection methods have had in this process. Technological developments like 
adaptive optics, space-based observatories like kepler and TESS, and high-precision spectrographs have made 
these approaches far more useful \cite{woodrum2023modest}. The discovery of smaller and farther-off planets has been made feasible by 
these developments. Exoplanet research is essential because it can increase our knowledge of planetary systems 
and the conditions necessary for habitability and the possibility that life exists elsewhere in the cosmos.

\subsection{Stars That Host Planets}
 On average, each star \cite{scalo2007m} is believed to host at least one planet. It's estimated that around 20\% of 
 sun like stars have an "earth-sized" planet in their habitable zones. The majority of known exoplanets 
 orbit stars similar to our Sun, specifically those in the main-sequence spectral categories F, G, or K \cite{lee2012detection} \cite{perryman2018exoplanet} \cite{meunier2023activity}. 
 Planets detectable by the radial-velocity method are less common around lower mass stars, such as the red dwarfs 
 in the M spectral group \cite{ananyeva2023exoplanets}. Despite this, the transit method, which excels at detecting smaller planets, 
 has enabled the kepler telescope to identify dozens of planets orbiting near red dwarfs.

  \begin{figure}
  \centering
\includegraphics[width=0.45\textwidth]{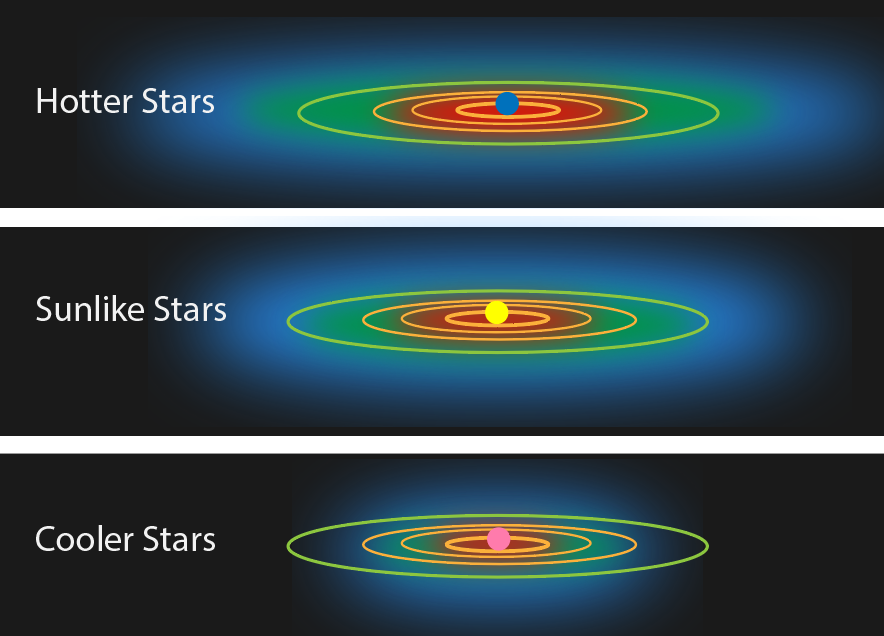}
  \caption{\scriptsize{Diagram showing the habitable zones surrounding certain stars. 
  The green area indicates where water on a planet's surface can stay liquid at the 
  ideal distance from the star. The blue area is too cool, and the red area is too heated.}}
  \label{fig:host stars}
\end{figure}
% \\

 \subsection{Detection methods}

Various detection methods \cite{rouan2023detection} exist for exoplanets, and astronomers continuously work 
toward developing new ways of detection. Some effective methods include radial velocity, 
which led to the discovery of the 1st exoplanet in the late '90s; the transit method, which 
analyzes light curves and periodograms \cite{bozza2016methods}; direct imaging, capturing exoplanets through images; 
gravitational microlensing; and astrometry \cite{perryman2018exoplanet}. Table 1 shows the total number of exoplanets 
discovered by various methods of detection till beginning of 2024 \cite{nasaexoplanet}. data taken from NASA 
exoplanet archive \cite{akeson2013nasa}. 
\\
 \begin{table}[htb]
    \tabularfont
 \centering
\caption{Exoplanets discovered by various methods till date}\label{discovered exoplanets}
\begin{tabular}{llllll}
 \topline 
%  \multicolumn{2}{|c|}{Discovered Exoplanets} \\
%  \hline
Method of discovery & Number of planets \\
 \midline
Astrometry&	3 \\
Disk Kinematics& 1\\
Radial Velocity&	1071\\
Eclipse timing variations&	17\\
Orbital brightness modulations&	9\\
Transit&	4146\\
Transit timing variations&	28\\
Pulsar timing variations&	7\\
Microlensing&	204\\
Imaging&	69\\
Pulsation timing variations&	2\\

 \hline
\end{tabular}
\end{table}

\subsection{Types of exoplanets} 
As of the conclusion of 2023, the existence of over 5500 planets had been confirmed. 
In an attempt to better understand these celestial bodies, scientists have undertaken 
numerous initiatives to categorize them based on various attributes, including their size, 
mass, and composition \cite{perryman2018exoplanet}.Table 2 shows the classification of exoplanets based on mass, size 
and chemical compositions of known gases such as $ZnS$, $H_2O$,
 $CO_2$ \& $CH_4$ \cite{forestano2023searching}.
 \begin{figure}
  \centering
  \includegraphics[width=0.5\textwidth,height=0.3997\textheight]{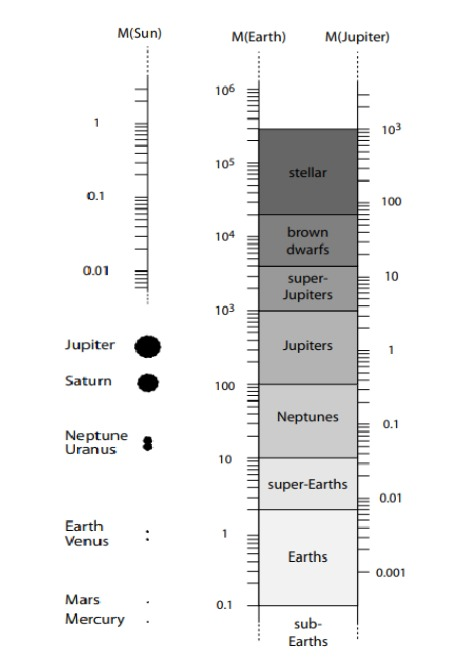}
  \caption{\scriptsize{planet masses are categorized based on solar system objects, as 
  suggested by Stevens \& Gaudi (2013). The solar system's planet masses are displayed on 
  the right, with the size of the circles corresponding to the planets' radii \cite{perryman2018exoplanet} \cite{mathur2021identifying}.}}
  \label{fig:exoplanets2}
\end{figure}

\begin{table}[htb]
\tabularfont
\centering
\caption{Types of exoplanets based on various factors}\label{my-label}
\begin{tabular}{llllllll}
\topline
Factor & Planet type & Parameter \\ 
\midline 
Size & Earth size & ($1.25-2R_\oplus$) \\
& Super Earth size & ($2-6R_\oplus$)\\
& Jupiter size & ($6-15R_\oplus$) \\
\\ \hline
 Mass & Earth & ($0.1M_\oplus-2M_\oplus$) \\
& Sub Earth & ($10^{-8}M_\oplus-0.1M_\oplus$)\\
& Super Earth & ($2M_\oplus-10M_\oplus$)\\
& Neptunes & ($10M_\oplus-100M_\oplus$)\\
& Jupiters & ($100M_\oplus-10^{3}M_\oplus$) \\
& Super Jupiters & ($10^{3}M_\oplus-13M_J$) \\
& Brown Dwarfs & ($13M_J-0.07M_\odot$)\\
& Stellar Companions &  ($0.07M_\odot-1M_\odot$)\\
\\ \hline
Composition  & Hot super Earths & ($187-1.12$) \\
(stellar flux) & Warm Super Earths & ($1.12-0.30$) \\
& Cold super Earth & ($0.30-0.0030$) \\
& Hot Rocky & ($182-1.0$) \\
& Warm Rocky & ($1.0-0.28$) \\
& Cold Rocky & ($0.28-0.0035$) \\
& Hot Sub Neptunes & ($188-1.15$) \\
& Warm Sub Neptunes & ($1.15-0.32$) \\
& Cold Sub Neptunes & ($0.32-0.0.0030$) \\
& Hot Sub Jovians & ($220-1.65$) \\
& Warm Sub Jovians & ($1.65-0.45$) \\
& Cold Sub Jovians & ($0.45-0.0030$) \\
& Hot Jovians  & ($220-1.65$) \\
& Warm Jovians & ($1.65-0.40$) \\
& Cold Jovians & ($0.40-0.0025$) \\
\\ \hline
\end{tabular}
\end{table}

 \subsection{Atmospheric conditions}
Finding transiting exoplanets is a significant breakthrough that provides previously 
unattainable knowledge about the composition and physical properties of these far-off worlds \cite{yakovlev2022exoplanet}. 
Transit and eclipse spectroscopy allows the investigation of atmospheric dynamics \cite{walker2018exoplanet} and heat 
transport processes, especially for hot Jupiters, while the resulting densities, derived from 
detailed measurements of masses and radii, offer important hints regarding internal structures 
and compositions. These discoveries have been made possible in large part by drawing on knowledge 
amassed over the previous 50 years from research on the planets and satellites of the solar system \cite{jakka2023assessing}.
  
\vspace{4px}
Future developments in photometric and spectroscopic observations, such as those brought about by 
the 2020 launch of the James Webb Space Telescope (JWST), are very promising. Advanced diagnostics 
may reveal the precise atmospheric compositions of exoplanets that are located in the "habitable zone \cite{jagtap2021habitability},
" which is where liquid water may be present. These compositions may include indications 
of either primitive or advanced life forms. These developments advance not only the hunt for 
extraterrestrial life but also help answer the important question of how common or uncommon life is 
in the universe \cite{seager2010exoplanet}. The dynamic interplay between theoretical models and observational methods is set 
to propel future research and transform our knowledge of exoplanetary systems.

\section{Exoplanet detection methods}
As discussed previously there are various detection methods which played an important role to 
present this unique field of study. all methods works on various different properties such as light, noise \cite{smith2008detection}, EMwaves etc. 
some of them are explained here.
\subsection{Radial Velocity Method}
The radial velocity technique for identifying exoplanets involves tracking the reflexive movement 
of a star, which is triggered by the gravitational force of a planet in orbit. This method 
assesses three visible characteristics of the star that are altered by the planet’s existence:\\

\textit{Radial Velocity}: When a planet orbits a star, both celestial bodies revolve around a 
shared center of mass, termed the barycenter. The star’s motion around this point results in a 
periodic fluctuation in its radial velocity \cite{hatzes2016radial}, which signifies its speed moving towards or away 
from an observer located on Earth. It is discovered by spectroscopic studies that this radial 
velocity shift exists. The spectral lines of a star change towards the blue end of the spectrum as 
it travels towards the observer, and to the red end when it goes away (blueshift and redshift, 
respectively) \cite{bozza2016methods} \cite{chen2023comparison}.\\ 

\textit{Angular (Astrometric) Position in the Sky}: The star moves in a modest circular or 
elliptical path around the shared center of mass due to the gravitational attraction of an orbiting 
planet \cite{pham2024new} \cite{bozza2016methods}. The star's angular location in the sky changes periodically, indicating its motion. 
Astrometric measures entail monitoring the star's apparent location with accuracy over time. Measurements 
of radial velocity have been used more frequently than this approach, which is more difficult \cite{perryman2018exoplanet} \cite{chen2023comparison}.\\ 

\textit{Time of Arrival of Periodic Reference Signal}: Variations in the arrival times of periodic 
signals, such pulsar pulses or eclipse timings in binary star systems, can be caused by the gravitational 
interaction between a star and its circling planet \cite{bozza2016methods}. Timing variations in these signals of reference 
may indicate the existence of an unseen planet and offer additional evidence of its impact \cite{fischer2015exoplanet}.\\ 

Radial velocity \cite{dumusque2017radial} measurements revealed the first credible discovery of an exoplanet in 1995 \cite{perryman2018exoplanet}. 
Since then, a number of exoplanets have been found using this method because to technological and 
observational breakthroughs. Via radial velocity metrics, 1075 exoplanets had been found as of 02/05/2024; 
some of these planets have been observed to transit their host stars, allowing the star's light to 
periodically fade as the planet passes in front of it.

% - - - - - - - - - - - - - - - - - - - - - - - - - - - - - - - - - - -
\subsection{Transit method}
Astronomy achieved a major milestone in 1999 with the ground-breaking observation of the first transit 
of an extrasolar planet \cite{perryman2018exoplanet} \cite{chen2023comparison}. The 'hot Jupiter' system HD 209458 was observed photometrically \cite{perryman2018exoplanet}, 
which allowed for this observation. A gas giant planet known as a "hot Jupiter" has a close orbit around 
its host star, which increases the planet's temperature somewhat. other star systems found by radial 
velocity surveys came into focus after this finding \cite{fischer2015exoplanet}. After a planet passes in front of its host 
star, causing a periodic dimming of the star's brightness, researchers kept an eye on these systems in 
the hopes of finding other transits \cite{yang2023detection}. Identifying exoplanets became increasingly dependent on this methodology, 
particularly for those that were not initially found via the radial velocity method \cite{bozza2016methods}. Dedicated surveys 
were rapidly launched to perform "blind hunts," initially from space then from ground-based observatories \cite{perryman2018exoplanet}. 
The goal of these surveys was to find new planets only by using the periodic transit fingerprints observed in 
star light curves. With this method, scientists were able to find a wide variety of exoplanets without any prior 
awareness of their presence \cite{chen2023comparison}. 
\begin{figure*}
  \centering
  \includegraphics[width=1\textwidth]{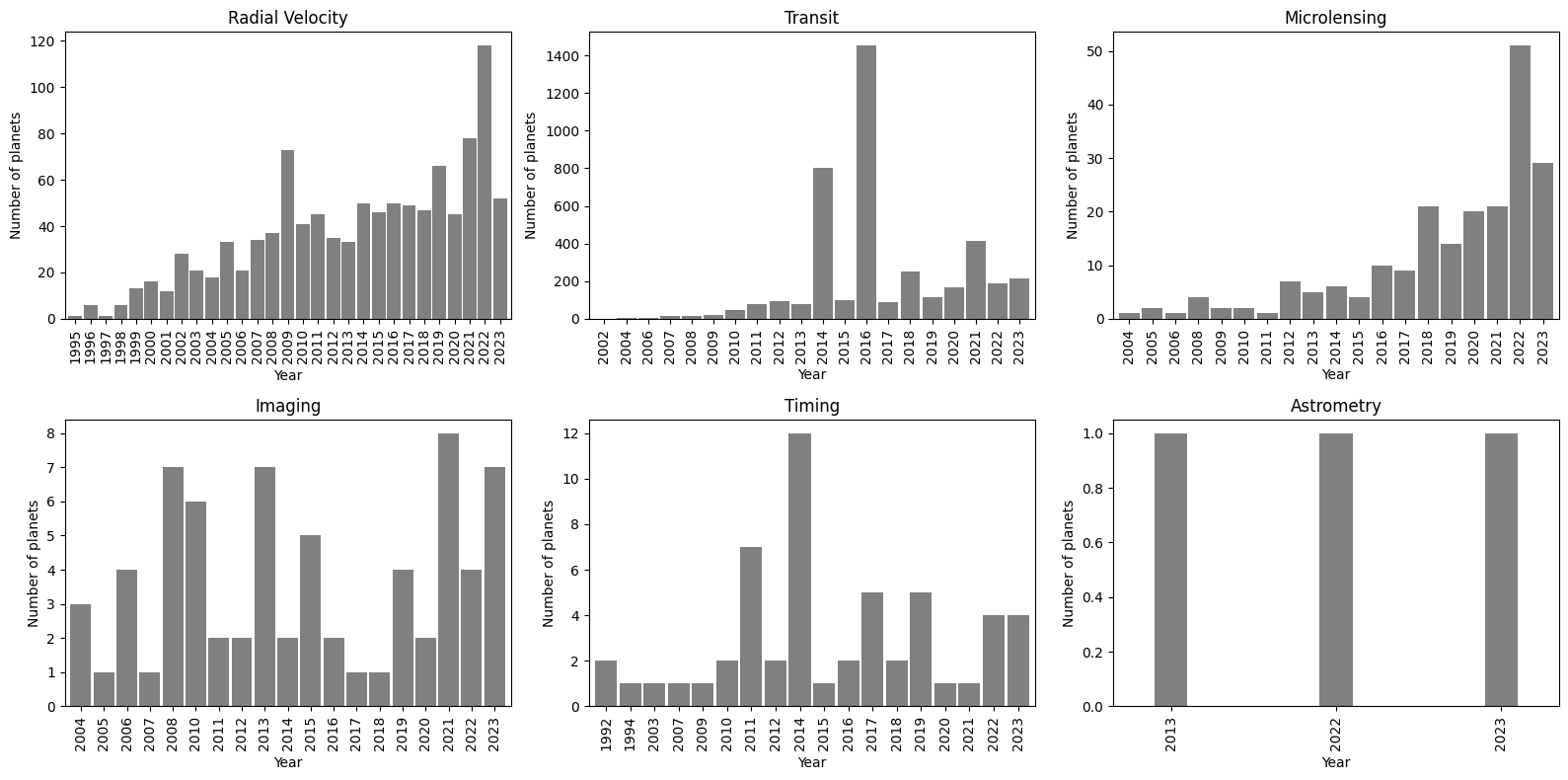}
  \caption{\scriptsize{Exoplanets detection methods. fig 4.1 represents Number of exoplanets detected 
  by Radial velocity method in previous years, fig 4.2 represents Number of exoplanets detected by Transit 
  method in previous years, fig 4.3 represents Number of exoplanets detected by Gravitational Microlensing 
  method in previous years, fig 4.4 represents Number of exoplanets detected by Direct Imaging method in 
  previous years, fig 4.5 represents Number of exoplanets detected by Timing method in previous years, 
  fig 4.6 represents Number of exoplanets detected by Astrometry method in previous years .}}
  \label{fig:exoplanets3}
\end{figure*}\\ 

In the field of exoplanet research, transiting planets are especially significant \cite{afanasev2018detection}. 
The light curves acquired during transits yield important details on planet sizes since the 
radii of the planets may be estimated due to the star's periodic dimming of light \cite{perryman2018exoplanet} \cite{fischer2015exoplanet}. 
Through techniques like radial velocity measurements, astronomers can establish the transiting 
planet's mass as well as its density, providing preliminary information about the planet's 
composition \cite{bozza2016methods} \cite{chen2023comparison}.
\begin{figure}
  \centering
  \includegraphics[width=0.5\textwidth]{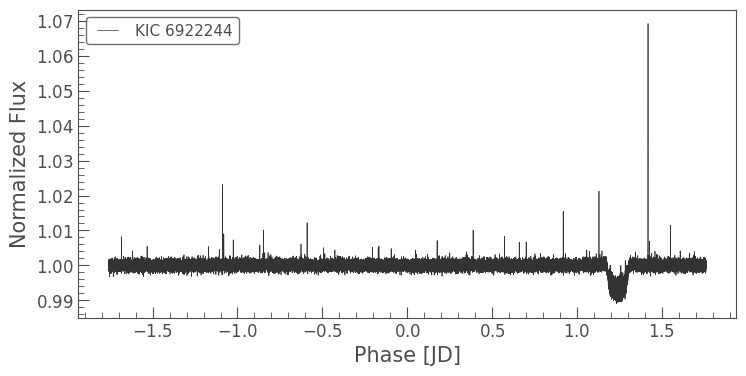}
  \caption{Representing the periodogram of exoplanet KIC 6922244}
  \label{fig:exoplanets4}
\end{figure}
\\Around the end of the year of 2023, exoplanet research had undergone a major transformation. 
There were now close to 4153 transiting planets known, with the Kepler space observatory 
detecting nearly 2778 of them as reported by NASA. Kepler began operations in 2009 and has been 
a key player in the area of transiting exoplanet research, providing an abundance of data that 
has made significant effects \cite{panahi2022detection} \cite{chen2023comparison}. A wide range of exoplanetary systems have been found as a 
result of the joint efforts of ground-based surveys \cite{perryman2018exoplanet}, such HAT and WASP, which revealed bright 
and easily detectable transiting systems, and Kepler's numerous discoveries, particularly those of 
close-orbiting multi-planet systems \cite{afanasev2018detection}. These systems contribute to our understanding of the 
diversity of planetary systems outside of our solar system by displaying a wide range of physical 
attributes and circumstances.

\subsection{Direct imaging method}
"Imaging" in the context of modern exoplanet research is the process of recognizing a planet as 
an individual source of light \cite{perryman2018exoplanet} \cite{chen2023comparison}. This may happen as a result of the planet's own thermal 
emission or light reflected from the host star. It is important that this imaging idea is not 
the same as reaching the planetary surface's spatial resolution \cite{bozza2016methods}. Additionally, it is distinct 
from techniques that detect fluctuations in the total brightness \cite{manfredi2020classification} of the planet and star system 
as a result of things like secondary eclipses, reflected light, or changes in the stellar spectrum \cite{birkby2018spectroscopic}
throughout a transit. In 2005, the first imaging detections were reported, namely the 
identification of huge, youthful, self-luminous planets in broad orbits \cite{bozza2016methods} \cite{perryman2018exoplanet}. The NASA exoplanet 
archive listed 44 of these discoveries as at the end of 2017 \cite{perryman2018exoplanet}. Prominent instances comprise 
planets discovered in well researched debris disk systems such as $\beta$ Pic and fomalhaut, in along 
with multi-planet systems like HR 8799 (which includes four planets) and LkCa 15 (containing two 
planets) \cite{perryman2018exoplanet}. There are certain deep learning models are used to inhance the capabilities of detection such as deep PACO \cite{flasseur2024deep}\\ 
Researchers have been able to determine the orbital velocity of planets in systems such as $\beta$ 
Pic, HR 8799, and fomalhaut thanks to multi-epoch research. Over the past ten years, developments 
in imaging instrument technology \cite{daglayan2023direct} \cite{hebrard2018exoplanet} have led to an emergence of cutting-edge devices specifically 
intended for exoplanet imaging \cite{beuzit2006direct}. VLT-SPHERE, Gemini-GPI \cite{galicher2014near}, and Subaru-HiCIAO are a few examples \cite{fischer2015exoplanet}. 
It is important to remember that these devices can only directly picture massive planets that are 
young, warm, and self-luminous at this time \cite{perryman2018exoplanet}. It is expected that with the next 30–40-meter 
telescopes, direct detection of point source pictures resulting from reflected starlight would 
become possible \cite{bozza2016methods} \cite{perryman2018exoplanet}. On the other hand, picturing planets similar to earth is still a distant 
objective \cite{chen2023comparison}. It is crucial to take into account imaging \cite{koechlin2005high} at radio and X-ray wavelengths while 
studying star-planet systems. These other wavelength ranges may provide light on many facets of 
the interactions that take place between stars and their planets \cite{flasseur2023combining}. There is talk about the potential 
to expand imaging capabilities to X-ray and radio wavelengths, which implies a more thorough 
investigation of observational methods in exoplanet studies \cite{rosa2023exoplanets} \cite{daglayan2023direct}.

% - - - - - - - - - - - - - - - - - - - - - - - - - - - - - - - - - - -
\subsection{Gravitational Microlensing}
The general relativity-based observation method known as gravitational microlensing \cite{perryman2018exoplanet}. 
Using gravitational microlensing, scientists found 210 exoplanets by the end of 2023. 
This technique uses the gravitational potential of a foreground object (the lens) to 
bend light beams from a distant background object (the source) \cite{bozza2016methods} \cite{perryman2018exoplanet}. Depending on how 
the source, lens, and observer are aligned, the outcome is a distorted image of the source, 
sometimes in many images \cite{chen2023comparison} \cite{beaulieu2023hunting}. The effects of gravitational lensing can be observed, 
leading to the classification into various domains \cite{jones2008exoplanets}. Robust lensing, observable at the 
level of individual objects, is further classified into macro- and microlensing. Whereas 
microlensing deals with discrete, unresolved multiple visuals, macrolensing generates multiple 
resolved images, or "arcs." Weak lensing is a different regime that is statistically noticed 
without a clear resolution of individual objects. Specific circumstances within gravitational 
lensing are described by the introduction of specialized words. Mesolensing is the term used to 
characterize the astrometric displacement that accompanied a high-proper motion foreground star, 
whereas nanolensing is the lensing caused by planetary-mass objects \cite{gaudi2012microlensing}.
Relative motion between the source, lens, and observer causes time-varying amplification of 
pictures . \\ 
Depending on the characteristics of the radiation source and the lens, this amplification 
takes place across an assortment of intervals, from hours to years. A more complex light curve 
may be seen from the background source if the foreground lensing object is gravitationally 
complex, such a star with planets or a cluster of galaxies \cite{wright2012exoplanet}. This complexity results from 
the changing alignment geometry's time-varying magnification. Significant discoveries have been 
made using gravitational microlensing \cite{tamanini2019gravitational} in exoplanet exploration. In 2004, a 4MJ planet was first 
clearly detected at approximately 4 astronomical units (au) \cite{perryman2018exoplanet}. A 5M $\oplus$  planet was then found 
in 2006. An observation of a two-planet system in 2008 made it possible to quantify orbital 
motion during the lensing event. In 2015, the lens mass and microlens parallax were measured. 
In early 2018, evidence for planets flying freely was proposed. These results validate 
gravitational microlensing as an independent and potent probe of exoplanets, especially over an 
important mass and orbital radius region \cite{bozza2016methods}.

\subsection{Astrometry}
\begin{table*}
\tabularfont
  \centering
 \caption{Comparative analysis of methods} 
 \begin{tabular}{p{2.5cm}|p{2cm}|p{2cm}|p{2cm}|p{2.5cm}|p{2.5cm}|p{2cm}}
  \topline
  Detection method&Principle&Measurements&Advantages&speciality&Drawbacks&Equipement\\
  \midline
Radial velocity&Gravitational Wobble&$m_p, T, a$&large amount of planets can be observe at once, precision&large planets, small cool stars&its for only planets closer to observer&Spectrometer\\
Transit photometry&Light Dip&$r_p, T, a$&High senstivity, Large sample&precise evaluation of properties, short period planets&False positives,intersecting and sort period planets&Photometer, TESS, Kepler, CoRoT\\
Direct Imaging&Visual Observation&$m_p, T, a$&characterization , atmospheres&large size and a, hot, planet close to observer&needed stability, glare&cornograph and infrared telescope\\
Gravitational Microlensing&Gravitational Lensing& $m_p$& Distant planets, galactic scale&background towards center of galaxy, large a&doubtful alignment, single trial&Warsaw telescope, OGLE\\
Astrometry&Stellar Wobble&RA, Dec,..&High precision, Long baselines&precise&extremely hard&Gaia, keck\\
  \hline
 \end{tabular}
 \end{table*}

%  \begin{table*}[htb]
%     \tabularfont
%     \caption{Caption text here}\label{secondTable}
%     \begin{tabular}{lccccccccccccr}
%     \topline
%     \textbf{head1}&\multicolumn{11}{c}{\textbf{head2}}&\textbf{head3}\\
%     \midline
%     one& two &three&four&five&six&seven&eight&nine&ten&eleven&twelve&thirteen\\
%     1&2&3&4&5&6&7&8&9&10&11&12&13\\
%     aaa&bbbb&cccc&ddddd&eee&ffff&ggggg&hhhhhhhh&iiii&kkkkkk&hhh&jjjjjj&lllll\\
%     \hline
%     \end{tabular}
%     \tablenotes{Table footnote here. Table spanning both the columns.}
%     \end{table*}

The exact measurement of the motions and positions of different celestial objects, from bodies 
of the solar system and stars within our galaxy to, theoretically, entire galaxies and clusters, 
is known as astrophysicist. Determining stellar parallaxes and proper movements has historically 
been a major goal of astrometry, as these are essential for comprehending the characteristics of 
host stars \cite{perryman2018exoplanet}.
In the present development, astrometry plays a new role: it seeks to identify the crosswise 
element of the displacement of a host star due to an orbiting planet's gravity. Measures of 
radial velocity, which are sensitive to the concomitant movement of the photocenter along the 
line-of-sight, are closely associated with this dynamic manifestation in the sky \cite{saffe2005ages}.\\

It has been difficult to achieve high-accuracy astrometry; hipparcos and the fine guidance 
sensors on the hubble space telescope, for example, achieved accuracies of about 1 milliarcsecond 
(mas). Still, this degree of accuracy has only just touched the tip of the iceberg in terms of how 
much star position displacements caused by planets in orbit can be identified. By the end of 2023, 
astrometry had only found three exoplanets, two of which were 5MJ and 2.26MJ in mass, and one of 
which was a substellar partner with a mass 28 times that of jupiter (28MJ). These discoveries 
were recorded in the NASA exoplanet archive. Since the gaia \cite{ranalli2018astrometry} mission was launched in 2013, the 
field of astrophysicists has seen a dramatic change \cite{panahi2022detection}. Over a billion stars with magnitudes 
up to V = 20–21 are being actively measured by gaia, with accuracies of about 20–25 microarcseconds
(µas) for stars with a magnitude of 15 \cite{perryman2018exoplanet}. Our knowledge of planetary systems is predicted to be 
completely transformed by this astounding increase in precision \cite{hodgkin2021gaia} \cite{panahi2022detection}. Several thousand planetary 
systems with correct planet masses and absolute orbits free of the sin i ambiguity (the uncertainty 
surrounding the inclination angle of the planetary orbit) should be found thanks to gaia's large 
dataset \cite{panahi2022detection} \cite{perryman2018exoplanet}. Moreover, gaia will provide insightful data on planetary systems' co-planarity, 
illuminating the configuration and behaviour of several planets within a system \cite{wiki:exosearchprojects}.

\section{Comparative analysis of detection methods}

Count of detected exoplanets by specific method is the proof of its importance. 
Every method of extrasolar planet detection works on different principles and scientific 
concepts \cite{perryman2018exoplanet} \cite{chen2023comparison}. While transit photometry and radial velocity methods are playing main role 
for the majority detection. after discovery there are various parameters and properties to 
know for detecting habitable planet, which is possible through other methods \cite{meunier2012comparison}. hence every 
detection method have its own advantages and disadvantages \cite{fischer2015exoplanet}. The transit photometry method 
coveres wide range of stars but its false positive rate is high \cite{tembhare2023amelioration}. transit method works for 
star-planet systems, which have orbits aligned in such a way that, they can be visible from earth. 
those exoplanets should move in between host star and earth. Radial velocity method is highly 
prolific and works for many planets but they must be close to the spectator \cite{mcarthur2004detection}. Timing method 
works for limited materiality, but provide accurate data of exoplanets residing far from earth. 
Direct imaging also works for few stars but straight forward detection of exoplanets. Large orbital 
radii planets can be detected by gravitational microlensing, but here measurements alignments can't 
be repeated and questionable \cite{hebrard2018exoplanet}. The majority of precise searches employ multiple methodologies 
to collect multifaceted information on a certain planet. In particular exoplanet HD 86266 c 
alternative name TOI-652 c detected by primary transit method in 2020 and gave radius parameter 
details. Table 3 shows the comparative analysis of detection methods with important parameters. 

\section{Space missions for exoplanets }

There are several missions are working towards exoplanets and there discoveries. 
these missions are running by various space agencies like NASA, ESA, JAXA, SETI etc. 
some missions ended up working and some are currently doing good work. here is the details 
of some space missions for exoplanets. The landscape of exoplanet exploration has been significantly shaped by a series of 
pioneering space missions, each contributing unique insights into the vast diversity of 
planetary systems beyond our solar neighborhood \cite{galicher2014near}. Among these missions, TESS (Transiting 
Exoplanet Survey Satellite) \cite{kane2020science} and GAIA \cite{ranalli2018astrometry} stand as ongoing endeavors, 

% \newpage
 \begin{table}[htb]
    \tabularfont
 \centering
\caption{Space missions for exoplanet detection} 
\begin{tabular}{p{4cm}|p{3.5cm}}
 \topline
%  \multicolumn{2}{|c|}{Space missions for exoplanets} \\
%  \hline
 Mission& Status \\
 \midline
TESS&	Ongoing \\
GAIA&	Ongoing\\
Kepler&	completed (2013)\\
K2&	Completed (2018)\\
CHEOPS&	On going\\
JWST&	On going\\
ARIEL&	Planned\\
RST&	Planned\\
TOLIMAN&	Planned\\
PLATO&	Planned\\
CoRot&	Completed(2006)\\
SPitzer&	Completed (2020)\\
 \hline
\end{tabular}
\end{table}

\vspace{4px}
tirelessly scanning the celestial expanse for transiting exoplanets and providing precise astrometric data for over a 
billion stars, respectively.There are various machine learning techniques \cite{barboza2020classifying} 
are used to study and detect exoplanets \cite{jin2022identifying}. Table 4 shows the list of missions 
working on exoplanet \cite{wiki:exosearchprojects} .

\vspace{4px}
Kepler, although completed in 2013, left an enduring legacy by unveiling thousands of 
exoplanets through the meticulous scrutiny of stellar light curves. Its successor, K2, 
continued this mission, expanding our understanding of exoplanets until its completion in 2018. 
Ongoing missions such as CHEOPS (Characterizing Exoplanet Satellite) and JWST (James Webb Space 
Telescope) \cite{alderson2023early} hold promise for further revelations \cite{agrawal2015comparative}. CHEOPS focuses on characterizing known 
exoplanets, scrutinizing their atmospheres and compositions, while JWST, a joint venture by NASA, 
ESA, and CSA, is poised to revolutionize infrared astronomy and deepen our understanding of 
exoplanet atmospheres \cite{hebrard2018exoplanet}.

\vspace{4px}
Looking ahead, a new wave of planned missions, including ARIEL (Atmospheric Remote-sensing 
Infrared Exoplanet Large-survey), RST (Roman Space Telescope), TOLIMAN, PLATO (PLAnetary Transits 
and Oscillations of stars), and more, signal a future of expanded horizons and enhanced 
capabilities in exoplanet detection and characterization. Reflecting on the completed missions, 
CoRot and Spitzer played pivotal roles, with CoRot contributing to the discovery of rocky 
exoplanets and Spitzer providing invaluable infrared observations until its completion in 2020. 
Together, these missions have propelled exoplanet science into a dynamic era, where each 
undertaking adds layers to the evolving narrative of the cosmos beyond our solar system \cite{wiki:exosearchprojects}.

\section{Conclusion and future scope}

A wide range of creative techniques used by astronomers and researchers have allowed the science 
of exoplanet detection to advance and diversify quite significantly. this detailed analysis of 
various exoplanet detection methods has illuminated the diverse approaches and technological 
advancements that have shaped our understanding of distant planetary systems. Each method, from 
the radial velocity and transit techniques to direct imaging, gravitational microlensing, 
astrometry, and pulsar timing, contributes uniquely to the growing catalog of known exoplanets. 
The precision of radial velocity measurements, the sensitivity of transits, and the potential for 
direct characterization through imaging collectively showcase the robustness of our exoplanet 
detection capabilities. Astrometry, with its emphasis on high precision and long baselines, offers 
complementary insights, especially for massive planets in close orbits.

\vspace{4px}
Future work in exoplanet detection should strive towards enhancing the synergy between 
these methods. Integrating data from multiple detection techniques will not only strengthen 
the validity of individual findings but also provide a more comprehensive view of exoplanetary 
systems. Additionally, advancements in technology, particularly in the development of 
next-generation space telescopes and ground-based observatories, hold the promise of further 
refining existing methods and unveiling new avenues for exoplanet discovery. Continued 
collaboration between astronomers, astrophysicists, and instrument engineers will be crucial for 
pushing the boundaries of exoplanet science, ultimately enriching our understanding of the vast 
diversity of planetary systems beyond our solar neighborhood.

%%References section
\bibliographystyle{apj}
\bibliography{main}

\begin{thebibliography}{}
\expandafter\ifx\csname natexlab\endcsname\relax\def\natexlab#1{#1}\fi

\bibitem[{Afanasev(2018)}]{afanasev2018detection}
Afanasev, D. 2018, arXiv preprint arXiv:1803.05565

\bibitem[{Agrawal {$et~al$.}(2015)Agrawal, Basak, Saha, Rosario-Franco, Routh,
  Bora, \& Theophilus}]{agrawal2015comparative}
Agrawal, S., Basak, S., Saha, S., {$et~al$.} 2015, researchgate

\bibitem[{Akeson {$et~al$.}(2013)Akeson, Chen, Ciardi, Crane, Good, Harbut,
  Jackson, Kane, Laity, Leifer, {$et~al$.}}]{akeson2013nasa}
Akeson, R., Chen, X., Ciardi, D., {$et~al$.} 2013, Publications of the
  Astronomical Society of the Pacific, 125, 989

\bibitem[{Alderson {$et~al$.}(2023)Alderson, Wakeford, Alam, Batalha,
  Lothringer, Adams~Redai, Barat, Brande, Damiano, Daylan,
  {$et~al$.}}]{alderson2023early}
Alderson, L., Wakeford, H.~R., Alam, M.~K., {$et~al$.} 2023, Nature, 614, 664

\bibitem[{Ananyeva {$et~al$.}(2023)Ananyeva, Ivanova, Shashkova, Yakovlev,
  Tavrov, Korablev, \& Bertaux}]{ananyeva2023exoplanets}
Ananyeva, V., Ivanova, A., Shashkova, I., {$et~al$.} 2023, Atmosphere, 14, 353

\bibitem[{Bai {$et~al$.}(2023)Bai, Lu, \& Orlofsky}]{bai2023dark}
Bai, Y., Lu, S., \& Orlofsky, N. 2023, Physical Review D, 108, 103026

\bibitem[{Barboza {$et~al$.}(2020)Barboza, Ulmer-Moll, \&
  Faria}]{barboza2020classifying}
Barboza, A., Ulmer-Moll, S., \& Faria, J. 2020, Classifying Exoplanets with
  Machine Learning, Tech. rep., Copernicus Meetings

\bibitem[{Beaulieu(2023)}]{beaulieu2023hunting}
Beaulieu, J.-P. 2023, Comptes Rendus. Physique, 24, 1

\bibitem[{Beuzit {$et~al$.}(2006)Beuzit, Mouillet, Oppenheimer, \&
  Monnier}]{beuzit2006direct}
Beuzit, J.-L., Mouillet, D., Oppenheimer, B.~R., \& Monnier, J.~D. 2006, HAL
  open science

\bibitem[{Birkby(2018)}]{birkby2018spectroscopic}
Birkby, J. 2018, Springer

\bibitem[{Bozza {$et~al$.}(2016)Bozza, Mancini, Sozzetti,
  {$et~al$.}}]{bozza2016methods}
Bozza, V., Mancini, L., Sozzetti, A., {$et~al$.} 2016, Astrophysics and Space
  Science Library: Berlin, Germany, 428

\bibitem[{Chen(2023)}]{chen2023comparison}
Chen, W. 2023, Highlights in Science, Engineering and Technology, 38, 235

\bibitem[{Daglayan {$et~al$.}(2023)Daglayan, Vary, Leplat, Gillis, \&
  Absil}]{daglayan2023direct}
Daglayan, H., Vary, S., Leplat, V., Gillis, N., \& Absil, P.-A. 2023, arXiv
  preprint arXiv:2304.03619

\bibitem[{Dumusque {$et~al$.}(2017)Dumusque, Borsa, Damasso, Diaz, Gregory,
  Hara, Hatzes, Rajpaul, Tuomi, Aigrain, {$et~al$.}}]{dumusque2017radial}
Dumusque, X., Borsa, F., Damasso, M., {$et~al$.} 2017, Astronomy \&
  Astrophysics, 598, A133

\bibitem[{Fischer {$et~al$.}(2015)Fischer, Howard, Laughlin, Macintosh,
  Mahadevan, Sahlmann, \& Yee}]{fischer2015exoplanet}
Fischer, D.~A., Howard, A.~W., Laughlin, G.~P., {$et~al$.} 2015, arXiv preprint
  arXiv:1505.06869

\bibitem[{Flasseur {$et~al$.}(2024)Flasseur, Bodrito, Mairal, Ponce, Langlois,
  \& Lagrange}]{flasseur2024deep}
Flasseur, O., Bodrito, T., Mairal, J., {$et~al$.} 2024, Monthly Notices of the
  Royal Astronomical Society, 527, 1534

\bibitem[{Flasseur {$et~al$.}(2023)Flasseur, Bodrito, Mairal, Ponce, Langlois,
  \& Lagrangev}]{flasseur2023combining}
Flasseur, O., Bodrito, T., Mairal, J., {$et~al$.} 2023, in 2023 31st European
  Signal Processing Conference (EUSIPCO), IEEE, 1723--1727

\bibitem[{Forestano {$et~al$.}(2023)Forestano, Matchev, Matcheva, \&
  Unlu}]{forestano2023searching}
Forestano, R.~T., Matchev, K.~T., Matcheva, K., \& Unlu, E.~B. 2023, The
  Astrophysical Journal, 958, 106

\bibitem[{Galicher {$et~al$.}(2014)Galicher, Rameau, Bonnefoy, Baudino, Currie,
  Boccaletti, Chauvin, Lagrange, \& Marois}]{galicher2014near}
Galicher, R., Rameau, J., Bonnefoy, M., {$et~al$.} 2014, Astronomy \&
  Astrophysics, 565, L4

\bibitem[{Gaudi(2012)}]{gaudi2012microlensing}
Gaudi, B.~S. 2012, Annual Review of Astronomy and Astrophysics, 50, 411

\bibitem[{Hatzes(2016)}]{hatzes2016radial}
Hatzes, A.~P. 2016, Methods of detecting exoplanets: 1st advanced school on
  exoplanetary science, 3

\bibitem[{Hebrard {$et~al$.}(2018)Hebrard, Belikov, Batalha, Mulders, Stark,
  Teal, Domagal-Goldman, Mandell, {$et~al$.}}]{hebrard2018exoplanet}
Hebrard, E., Belikov, R., Batalha, N.~M., {$et~al$.} 2018, arXiv preprint
  arXiv:1802.09602

\bibitem[{Hodgkin {$et~al$.}(2021)Hodgkin, Harrison, Breedt, Wevers, Rixon,
  Delgado, Yoldas, Kostrzewa-Rutkowska, van Leeuwen, Blagorodnova,
  {$et~al$.}}]{hodgkin2021gaia}
Hodgkin, S., Harrison, D., Breedt, E., {$et~al$.} 2021, Astronomy \&
  Astrophysics, 652, A76

\bibitem[{Jagtap {$et~al$.}(2021)Jagtap, Inamdar, Dere, Fatima, \&
  Shardoor}]{jagtap2021habitability}
Jagtap, R., Inamdar, U., Dere, S., Fatima, M., \& Shardoor, N.~B. 2021, in 2021
  IEEE International IOT, Electronics and Mechatronics Conference (IEMTRONICS),
  IEEE, 1--6

\bibitem[{Jakka(2023)}]{jakka2023assessing}
Jakka, M.~S. 2023, arXiv preprint arXiv:2305.11204

\bibitem[{Jin {$et~al$.}(2022)Jin, Yang, \& Chiang}]{jin2022identifying}
Jin, Y., Yang, L., \& Chiang, C.-E. 2022, arXiv preprint arXiv:2204.00721

\bibitem[{Jones(2008)}]{jones2008exoplanets}
Jones, B.~W. 2008, International Journal of Astrobiology, 7, 279

\bibitem[{Kane {$et~al$.}(2020)Kane, Bean, Campante, Dalba, Fetherolf, Mocnik,
  Ostberg, Pepper, Simpson, Turnbull, {$et~al$.}}]{kane2020science}
Kane, S.~R., Bean, J.~L., Campante, T.~L., {$et~al$.} 2020, Publications of the
  Astronomical Society of the Pacific, 133, 014402

\bibitem[{Koechlin {$et~al$.}(2005)Koechlin, Serre, \&
  Duchon}]{koechlin2005high}
Koechlin, L., Serre, D., \& Duchon, P. 2005, Astronomy \& Astrophysics, 443,
  709

\bibitem[{Kumari {$et~al$.}(2023)}]{kumari2023identification}
Kumari, A., {$et~al$.} 2023, arXiv preprint arXiv:2305.09596

\bibitem[{Lee {$et~al$.}(2012)Lee, Mkrtichian, Han, Park, \&
  Kim}]{lee2012detection}
Lee, B.-C., Mkrtichian, D., Han, I., Park, M.-G., \& Kim, K.-M. 2012, Astronomy
  \& Astrophysics, 548, A118

\bibitem[{Manfredi {$et~al$.}(2020)Manfredi, Areias, Pupo, Santos, \&
  Valio}]{manfredi2020classification}
Manfredi, G.~M., Areias, H.~S., Pupo, N.~L., Santos, I., \& Valio, A. 2020,
  Boletim da Sociedade Astron{\^o}mica Brasileira, 31, 21

\bibitem[{Mathur {$et~al$.}(2021)Mathur, Sizon, \&
  Goel}]{mathur2021identifying}
Mathur, S., Sizon, S., \& Goel, N. 2021, in Advances in Machine Learning and
  Computational Intelligence: Proceedings of ICMLCI 2019, Springer, 369--379

\bibitem[{McArthur {$et~al$.}(2004)McArthur, Endl, Cochran, Benedict, Fischer,
  Marcy, Butler, Naef, Mayor, Queloz, {$et~al$.}}]{mcarthur2004detection}
McArthur, B.~E., Endl, M., Cochran, W.~D., {$et~al$.} 2004, The Astrophysical
  Journal, 614, L81

\bibitem[{Meunier {$et~al$.}(2012)Meunier, Lagrange, \&
  De~Bondt}]{meunier2012comparison}
Meunier, N., Lagrange, A.-M., \& De~Bondt, K. 2012, Astronomy \& Astrophysics,
  545, A87

\bibitem[{Meunier {$et~al$.}(2023)Meunier, Pous, Sulis, Mary, \&
  Lagrange}]{meunier2023activity}
Meunier, N., Pous, R., Sulis, S., Mary, D., \& Lagrange, A.-M. 2023, Astronomy
  \& Astrophysics, 676, A82

\bibitem[{NASA(2024)}]{nasaexoplanet}
NASA. 2024, How we find and characterize exoplanets,
  \url{https://exoplanets.nasa.gov/discovery/how-we-find-and-characterize/}

\bibitem[{Panahi {$et~al$.}(2022)Panahi, Zucker, Clementini, Audard,
  Binnenfeld, Cusano, Evans, Gomel, Holl, Ilyin,
  {$et~al$.}}]{panahi2022detection}
Panahi, A., Zucker, S., Clementini, G., {$et~al$.} 2022, Astronomy \&
  Astrophysics, 663, A101

\bibitem[{Perryman(2018)}]{perryman2018exoplanet}
Perryman, M. 2018, The exoplanet handbook (Cambridge university press)

\bibitem[{Pham {$et~al$.}(2024)Pham, Rein, \& Spiegel}]{pham2024new}
Pham, D., Rein, H., \& Spiegel, D.~S. 2024, arXiv preprint arXiv:2401.02849

\bibitem[{Ranalli {$et~al$.}(2018)Ranalli, Hobbs, \&
  Lindegren}]{ranalli2018astrometry}
Ranalli, P., Hobbs, D., \& Lindegren, L. 2018, Astronomy \& Astrophysics, 614,
  A30

\bibitem[{Rosa \& do~Nascimento~Dias(2023)}]{rosa2023exoplanets}
Rosa, L.~N., \& do~Nascimento~Dias, B. 2023, Revista Mexicana de Astronom{\i}a
  y Astrof{\i}sica Serie de Conferencias (RMxAC), 55, 137

\bibitem[{Rouan \& Lagrange(2023)}]{rouan2023detection}
Rouan, D., \& Lagrange, A.-M. 2023, Comptes Rendus. Physique, 24, 1

\bibitem[{Saffe {$et~al$.}(2005)Saffe, G{\'o}mez, \& Chavero}]{saffe2005ages}
Saffe, C., G{\'o}mez, M., \& Chavero, C. 2005, Astronomy \& Astrophysics, 443,
  609

\bibitem[{Scalo {$et~al$.}(2007)Scalo, Kaltenegger, Segura, Fridlund, Ribas,
  Kulikov, Grenfell, Rauer, Odert, Leitzinger, {$et~al$.}}]{scalo2007m}
Scalo, J., Kaltenegger, L., Segura, A., {$et~al$.} 2007, Astrobiology, 7, 85

\bibitem[{Seager \& Deming(2010)}]{seager2010exoplanet}
Seager, S., \& Deming, D. 2010, Annual Review of Astronomy and Astrophysics,
  48, 631

\bibitem[{Shafi(2021)}]{mediumshafidetection1}
Shafi. 2021, Exoplanets detection using Artificial Intelligence (AI),
  \url{https://medium.datadriveninvestor.com/exoplanets-detection-using-artificial-intelligence-ai-a969316eb37/}

\bibitem[{Smith {$et~al$.}(2008)Smith, Ferrari, \&
  Carbillet}]{smith2008detection}
Smith, I., Ferrari, A., \& Carbillet, M. 2008, IEEE Transactions on Signal
  Processing, 57, 904

\bibitem[{Tamanini \& Danielski(2019)}]{tamanini2019gravitational}
Tamanini, N., \& Danielski, C. 2019, Nature Astronomy, 3, 858

\bibitem[{Tembhare {$et~al$.}(2023)Tembhare, Patil, Gadre, \&
  Pandit}]{tembhare2023amelioration}
Tembhare, S., Patil, A., Gadre, S., \& Pandit, S.~V. 2023, International
  Research Journal of Engineering and Technology

\bibitem[{Walker {$et~al$.}(2018)Walker, Bains, Cronin, DasSarma, Danielache,
  Domagal-Goldman, Kacar, Kiang, Lenardic, Reinhard,
  {$et~al$.}}]{walker2018exoplanet}
Walker, S.~I., Bains, W., Cronin, L., {$et~al$.} 2018, Astrobiology, 18, 779

\bibitem[{Wei(2018)}]{wei2018survey}
Wei, J. 2018, arXiv preprint arXiv:1805.02771

\bibitem[{wikipedia(2022)}]{wiki:exosearchprojects}
wikipedia. 2022, List of exoplanet search projects,
  \url{https://en.wikipedia.org/wiki/List_of_exoplanet_search_projects}

\bibitem[{Woodrum {$et~al$.}(2023)Woodrum, Hviding, Amaro, \&
  Chamberlain}]{woodrum2023modest}
Woodrum, C., Hviding, R.~E., Amaro, R.~C., \& Chamberlain, K. 2023, arXiv
  preprint arXiv:2303.16915

\bibitem[{Wright \& Gaudi(2012)}]{wright2012exoplanet}
Wright, J.~T., \& Gaudi, B.~S. 2012, arXiv preprint arXiv:1210.2471

\bibitem[{Yakovlev {$et~al$.}(2022)Yakovlev, Valeev, Valyavin, Tavrov, Aitov,
  Mitiani, Korablev, Galazutdinov, Beskin, Emelianov,
  {$et~al$.}}]{yakovlev2022exoplanet}
Yakovlev, O.~Y., Valeev, A.~F., Valyavin, G.~G., {$et~al$.} 2022, Frontiers in
  Astronomy and Space Sciences, 9, 903429

\bibitem[{Yang(2023)}]{yang2023detection}
Yang, G. 2023, Highlights in Science, Engineering and Technology, 31, 196

\end{thebibliography}
\end{document}